\documentclass[doublecol]{epl2} 
\usepackage{amsmath}
\usepackage{amsfonts}
\usepackage{amssymb}
\usepackage{graphicx}
\usepackage{color}
\usepackage{xcolor}
\usepackage{cite}
\usepackage{hyperref}

\newcommand{\beq}{\begin{equation}}
\newcommand{\eeq}{\end{equation}}
\newcommand{\bea}{\begin{eqnarray}}
\newcommand{\eea}{\end{eqnarray}}

\newcommand{\nn}{\nonumber}

\title{Slow scrambling in sonic black holes}
\shorttitle{Slow scrambling in sonic black holes} %Insert here a short version of the title if it exceeds 70 characters

\author{G. Menezes\inst{1,2} \and J. Marino\inst{3,4}}
\shortauthor{G. Menezes and J. Marino}

\institute{                    
  \inst{1} Department of Physics, University of Massachusetts, Amherst, Massachusetts 01003, USA\\
  \inst{2} Departamento de F\'isica, Universidade Fe\-de\-ral Rural do Rio de Janeiro, 23897-000 Serop\'edica, RJ, Brazil\\
  \inst{3} Department of Physics and Center for Theory of Quantum Matter,
University of Colorado Boulder, Boulder, Colorado 80309, USA\\
  \inst{4} JILA, NIST and University of Colorado,
Department of Physics, University of Colorado, Boulder, CO 80309, USA	 
}

\abstract{We study from the perspective of quantum information scrambling an acoustic black hole modeled by two semi-infinite, stationary, one dimensional condensates, connected by a spatial step-like discontinuity, and flowing respectively at subsonic and supersonic velocities. We develop a simple analytical treatment based on Bogolyubov theory of quantum fluctuations which is sufficient to derive analogue Hawking emission, and we compute out-of-time order correlations (OTOCs) of the Bose density field. We find that a large class of sonic black holes presents slow scrambling contrary to their astrophysical counterparts. This manifests in a power law growth $\propto t^2$ of OTOCs in contrast  to the exponential increase in time expected for fast  scramblers.}

\begin{document}

\maketitle

%%%%%%%%%%%%%%%%%%%%
\section{Introduction}

There has been recently   a rise of activity and interest on the problem of quantum information scrambling conducted at the border of condensed matter physics and gravity~\footnote{A. Kitaev, Talks at KITP:

http://online.kitp.ucsb.edu/online/entangled15/kitaev/ 

http://online.kitp.ucsb.edu/online/entangled15/kitaev2/

April and May 2015}.
The connection originates from studies regarding infalling matter into  two-dimensional black holes,   which are capable of producing shock waves  propagating along the background metric and detectable only by an unconventional type of out-of-time order correlators (OTOCs) -- being the ordered correlation functions (e.g. in Keldysh field theory) insensitive to the gravitational disturbance~\cite{kitaev}.  These out-of-time order correlators, $F(\tau)$, are  generally defined as~\cite{larkin,shenker,shenker2,maldacena,faoro,galit}
\begin{equation}\label{eq:otoc}
F(\tau)=\langle \hat{W}^\dag(\tau)\hat{V}^\dag \hat{W}(\tau)\hat{V} \rangle,
\end{equation} 
{for two operators $\hat{W}$ and $\hat{V}$ which commute at initial time}, evolving under the Hamiltonian $H$ as $\hat{W}(\tau) \equiv e^{iH\tau}\hat{W}e^{-iH\tau}$.
OTOCs are currently considered a sensible measure of loss of memory of an initial state or of information scrambling in a quantum system:
$F(\tau)$ in Eq. \eqref{eq:otoc} measures the overlap between two states, the former obtained by applying the operator $\hat{V}$ at time $\tau=0$ and  $\hat{W}$ at a later time $\tau$, while the latter is engineered reversing this procedure in time -- namely applying first $\hat{W}$ at time $\tau$ and then applying the operator $\hat{V}$ back at the initial time $\tau=0$. $F(\tau)$  encodes as well the growth in time of the commutator of $\hat{W}$ and $\hat{V}$, due to interactions, via the relation 
\begin{equation}\label{relaz}
\langle|[\hat{W}(t),\hat{V}]|^2\rangle=2[1-\textrm{Re}(F(\tau))].
\end{equation}
%, where $G(\tau)$ is the ``time-ordered version" of $F(\tau)$.
%[\textcolor{red}{Please check this, because in the previous version the time-ordered term was not present and I think it should be present}]. 
%
~Its allure and its wide use in current research stems from several thrilling aspects.

In the problem of matter falling into the black hole, the outgoing shock wave leaves on the OTOC an exponentially growing signature at short times $\propto \exp{(\lambda t)}$, where $\lambda=2\pi T_{H}$ {and $T_{H}$ is the Hawking temperature of the black hole}. This fact qualifies black holes as the fastest scramblers in nature~\cite{susskind}, as a consequence of a bound, $\lambda\leq 2\pi T$, derived by Maldacena \emph{et al.}~\cite{maldacena} {on the   exponent, $\lambda$, regulating the growth of OTOCs ($T$ in this latter case is the temperature of the specific system under inspection}). The bound is hard to saturate for conventional condensed matter systems, although few of them~\cite{Das, ON, stanfordd, patel, bord} exhibit an exponentially linear increase of $F(\tau)$. Only  in a recent extension by Kitaev of an original model by Sachdev and Ye~\cite{SY, bagrets2} (consisting of pairwise coupled $SU(M)$ spins with  disordered interactions, see also~\cite{parcol}), is information  scrambled as fast as in a black hole. The model comprises several Majorana sites connected by a disordered four body interaction with variance $J$~\cite{altman2, gross}, and in the low temperature limit, $ J\gg T$, displays an OTOC~\cite{Maldacenareview, bagrets} growing with an exponent saturating the bound, $\lambda\simeq 2\pi T$. 

A recent semi-classical analysis of this Sachdev-Ye-Kitaev (SYK) model has shed light on the physical interpretation of the bound~\cite{scafidi}: in  a system with a hypotethical scrambling exponent $\lambda'$ larger than the maximal one of  the SYK model, quantum interference effects would  renormalize $\lambda'$ to a value smaller than $2\pi T$. Equally interesting is  the semi-classical limit, $F_C(\tau)$, of Eq.~\eqref{eq:otoc}: it was realized sometime ago that, in a one-particle quantum chaotic system, $F_C(\tau)\propto (\partial p(t)/\partial q(0))^2\sim e^{\lambda_L t}$,  for two canonically conjugated operators as $\hat{W}=\hat{q}$ and $\hat{V}=\hat{p}$~\cite{larkin}. This makes transparent the interpretation of the OTOC as a measure of the sensitivity  of particles' trajectories to initial conditions, because of the resemblance between the growth of $F_C(\tau)$ and  the traditional Lyapunov exponent, $\lambda_L$~\cite{gutz}. Primarily as a result of this analogy, out-of-time order correlators have been promoted  as quantifiers of quantum chaos in the context of information scrambling~\cite{hosur}.

On the other hand, slow scramblers are systems with out-of-time order correlations growing polynomially in time, and examples range from the isotropic (or anisotropic) Dicke model~\cite{dicke}, to interacting fermions on the lattice in infinite dimensions~\cite{tsuji}, Luttinger Liquids~\cite{moess}, quantum systems in the many-body localized phase~\cite{he,Huse,Fan}, as well as  integrable and non-integrable periodically kicked quantum Ising chains~\cite{kuk}.

Inspired by the fundamental connection between quantum chaos and black holes, in this work we regard the sonic dumb holes -- the condensed matter analog of astrophysical black holes~\cite{leon, garay, visser,horst, unruh2}, from the perspective of information scrambling, and we compute from the microscopic theory of Bogolyubov quantum fluctuations in Bose gases their out-of-time order correlations.

Sonic black holes were conceived  in an early work by Unruh~\cite{unruh}, who elucidated that sound waves propagation in an inhomogeneous fluid are ruled by the same dynamical equations of a massless scalar field  in a Schwarzschild spacetime. In particular, the simplest condensed matter analog of a dumb hole is  realized by a stationary  one-dimensional Bose-Einstein condensate with a boundary separating an upstream subsonic flowing region from a downstream supersonic one; at this `horizon' phonons are emitted out of the quantum vacuum in the form of correlated pairs traveling into the sub- and supersonic regions respectively~\cite{rec09,may11,balb08}. These pairs  leave a characteristic signal in the density-density correlation of the Bose gas for points located on opposite sites of the horizon, and, strikingly, the power spectrum of phonons emitted in the upstream is thermal at a temperature proportional to the inverse of the fundamental length scale of the Bose gas, the  healing length~\cite{rec09}. Such thermal emission represents the phononic version of the Hawking effect, while the sub-sonic downstream flow takes the role of the interior of the black hole, since any sound wave emitted would be dragged away without reaching the upstream flow.

Despite their precocious conception, sonic black holes have been only recently created in cold gases~\cite{expBH1}, and the reported measurement of analog Hawking radiation itself is an achievement of the last few years~\cite{expBH2, nuovostein}, although the complete interpretation of the observed phenomena is still in intensive dispute (see, for instance, Ref.~\cite{parola} and references cited therein).

%%%%%%%%%%%%%%%%%%%%
\section{The model}

Before discussing  technical aspects of the calculation of OTOCs in dumb  holes, we summarize the physics of sonic horizons, following~\cite{rec09, may11}, and discuss the main findings of such works.

\begin{figure}
 \centering
    \includegraphics[width=8.8cm]{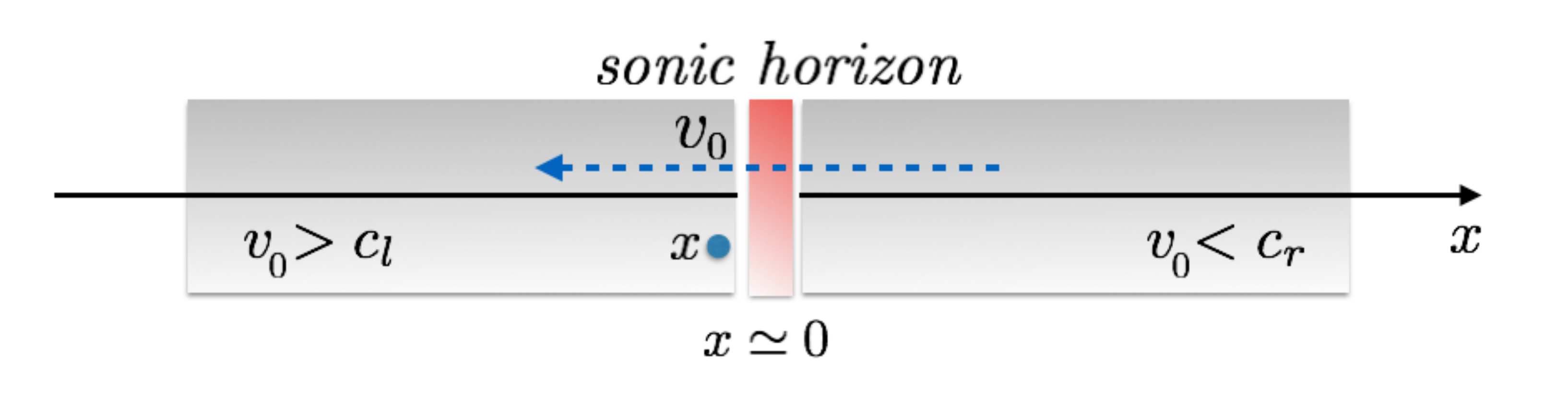}
    \caption{Schematic representation of the sonic black hole: the velocity of the flow is supersonic ($v_{0}>c_l$) in the downstream region, $x<0$, and subsonic ($v_{0}<c_r$) in the upstream one, $x>0$. This occurs as a consequence of a sonic horizon imposed by an external potential around $x\simeq0$; at such point, correlated pairs of particles are emitted and an analogue Hawking effect is realised~\cite{rec09,may11}. We compute the out-of-time order correlator of the Bose density field, Eq.~\eqref{corr_comm_L}, inside the acoustic hole, at a point $x$ (blue dot in figure)  close to the sonic horizon.
 }
     \label{fig}
\end{figure}

We consider a one-dimensional interacting Bose gas with mass, $m$, and with a space-dependent interaction coupling, $g(x)$, whose dynamics is governed by the Gross-Pitaevskii equation. The flow of the gas is kept stationary by an external potential, $V_{ext}(x)$, and at the spatial point $x=0$, its velocity  encounters an abrupt variation. We assume the condensate to have uniform density $\rho_0$ and to flow with a uniform velocity $v_{0}$ along the negative $x$-axis. The external potential $V_{\textrm{ext}}(x)$ and the repulsive atom-atom interaction coupling $g(x)$ are taken  homogeneous within each region, and satisfying the condition $V^{l}_{\textrm{ext}} + g^{l} \rho_0 = V^{r}_{\textrm{ext}} + g^{r} \rho_0$. The superscripts $l$ and $r$ refer respectively to the left ($x <0$)  and right ($x >0$) halves of the wire,  with a discontinuity located at $x = 0$. The  abrupt change in $g$ is at the origin of the sudden change in the local  sound speed, from $c_{l}$ to $c_{r}$, as it can be directly realised from the standard relation $m c^2 = \rho_0 g$.  As we mentioned above,  when the downstream region is supersonic $v_{0}>c_l$ and the upstream region flows subsonically $v_{0}<c_r$, a dumb-hole is realized, impeding a sound wave emitted in the downstream region to reach the upstream one, in an analogous fashion to trapping of light and matter in the interior of a black hole.

Fig.~\ref{fig} outlines the physics of acoustic black holes studied in this work.
In the dilute regime, we can use Bogolyubov theory~\cite{peth,pita} to describe small quantum fluctuations, $\delta\rho(x)$, on the top of the one-dimensional,  condensate density
$\rho_0$ through $\rho(x)\simeq \rho_0+\delta\rho(x)$, where the density operator $\rho(x)\equiv\psi^\dag(x)\psi(x)$ is expressed in terms of the field amplitude $\psi(x)$. The latter can be equivalently expanded in small fluctuations as follows: $\psi(t,x)\sim\psi_0(x)[1+\delta\psi(t,x)]\,\exp{(-i \mu t)}$, where $\mu$ is the frequency oscillation of the ground state wavefunction. The density operator is written therefore in the form
\beq
\rho(x) = |\psi_{0}(x)|^{2}+\delta \rho(x,t),
\eeq
where one has $\rho_0=|\psi_0(x)|^2$ and  $\delta\rho(x,t)\simeq|\psi_{0}(x)|^{2}(\delta\psi(t,x)+\delta\psi^{\dagger}(t,x))$. The  operator describing fluctuations of the Bose amplitude $\delta\psi(t,x)$, satisfies the Bogolyubov-de Gennes equations and the momentum-resolved Bogolyubov excitation spectrum reads in this case (we fix $\hbar = 1$)
\beq
\omega^{\pm}_k  = v_{0} k \pm c_\alpha k\sqrt{1 + \left(\frac{k\xi_\alpha}{2}\right)^2},
\label{dispersion}
\eeq
where the upper (lower) sign corresponds to the positive (negative) norm Bogolyubov branches~\cite{peth,pita}, while the index $\alpha=l,r$, discriminates velocities and healing lengths on opposite sides of the sonic black hole. The dispersion relation~\eqref{dispersion} is the key to understand the physics of  Hawking emission in the analogue black hole. There also exists a threshold frequency $\Omega$ for Bogolyubov excitations, as a consequence of the supersonic character of the flow in the downstream region~\cite{rec09, may11}, below which the analog Hawking emission of the  dumb hole are manifest in density correlation functions. This threshold frequency is given by the maximum frequency of the negative norm Bogolyubov mode in the downstream supersonic region; it corresponds to the Bogolyubov frequency with mode  $k_{*}$, obtained from the extremal values of the dispersion relation, $d\omega^{-}_k/dk = 0$. 

For $0 < \omega < \Omega$, four real roots of the dispersion relation exist in the supersonic flowing region, corresponding to four oscillatory modes, two on the positive norm branch and the other two on the negative norm one. Two of these modes, named $k_{1}$ and $k_{2}$, lie in the small momentum region and have  hydrodynamic character. Both of them propagate with a negative group velocity. The $k_1$ solution belongs to the positive norm branch, whereas the $k_2$ solution belongs  to the negative norm one and the corresponding excitation quanta carry a negative energy. On the other hand, the wavevectors associated with the other two roots, indicated as $k_3$ and $k_4$,  respectively in positive and negative norm branches, are non-perturbative in $\xi$, and hence lie well outside the hydrodynamic region. In addition, they propagate in the positive direction along the upstream region. 

The physical picture that emerges is that pairs of phonons are continuously generated by the horizon and then one propagates within the subsonic region and the other to the supersonic part, at speeds $v_{0}+c_r$ (positive) and $v_{0}+c_l$ (negative) respectively, and correlating the Bose density field  at positions $z=(v_{0}+c_r)t$ and $z'=(v_{0}+c_l)t$, at the time $t$ after emission~\cite{rec09, may11}. For excitation energies larger than $\Omega$, the modes $k_3$ and $k_4$ become complex, corresponding to decaying and growing modes, causing the disappearance of the Hawking signal in the density-density correlation of the Bose gas.

We now briefly outline the technical procedure employed in our calculations, which follows closely the  method presented in~\cite{may11} and reviewed in Ref.~\cite{book1}. The starting point consists in expanding the quantum amplitude of the Bose gas in the basis of the incoming modes
\bea\label{eq:expansion}
\delta\psi(t,x) &=& \int_{0}^{\Omega}\,d\omega \Bigl[a^{v, in}_{\omega} \phi^{in}_{v,r}(t,x) 
+ a^{3, in}_{\omega} \phi^{in}_{3,l}(t,x) 
\nn\\
&+&\, (a^{4, in}_{\omega})^{\dagger} \phi^{in}_{4,l}(t,x) 
+ (a^{v, in}_{\omega})^{\dagger}\tilde{\phi}^{in\,*}_{v,r}(t,x) 
\nn\\
&+&\, (a^{3, in}_{\omega})^{\dagger} \tilde{\phi}^{in\,*}_{3,l}(t,x) + a^{4, in}_{\omega}\tilde{\phi}^{in\,*}_{4,l}(t,x)\Bigr],
\eea
where the notation $\phi, \tilde{\phi}$ is employed in order to distinguish between different mode normalizations. In the above expression, the labels $l$ and $r$ refer to superposition of plane waves, respectively propagating in the downstream (left), $x<0$, and upstream (right), $x>0$, regions, and evolving with the Bogolyubov eigenenergies~\eqref{dispersion}; explicit expressions can be found in Ref.~\cite{may11}. In the downstream region the modes $k_3$ and $k_4$ are the incoming ones, while in the upstream, subsonic region the mode labelled as $v$ in Eq.~\eqref{eq:expansion} describes the propagation of waves towards the horizon placed at $x=0$. The bosonic annihilation operators in Eq.~\eqref{eq:expansion} define a vacuum state,
\beq\label{vac}
a^{v, in}_{\omega}|0,in\rangle = 0,\,\,\, a^{u, in}_{\omega}|0,in\rangle = 0 
\eeq
with $u=3,4$ labelling modes in the downstream region. In turn, the expansion analogous to~\eqref{eq:expansion} in terms of outgoing wave-packets is given by 
\bea
\hspace{-4mm}
\delta\psi(t,x) &=& \int_{0}^{\Omega}\,d\omega \Bigl[a^{v, out}_{\omega} \phi^{out}_{v,l}(t,x) 
+ a^{u r, out}_{\omega} \phi^{out}_{u,r}(t,x) 
\nn\\
&+&\, (a^{u l, out}_{\omega})^{\dagger} \phi^{out}_{u,l}(t,x) 
+ (a^{v, out}_{\omega})^{\dagger}\tilde{\phi}^{out\,*}_{v,l}(t,x) 
\nn\\
&+&\, (a^{u r, out}_{\omega})^{\dagger} \tilde{\phi}^{out\,*}_{u,r}(t,x) + a^{u l, out}_{\omega}
\tilde{\phi}^{out\,*}_{u,l}(t,x)\Bigr]
\label{eq:expansion2}
\eea
and again an expression for such modes can be found in Ref.~\cite{may11}. Here in the downstream region the modes $k_1$ and $k_2$ are the outgoing ones, and in the upstream region the mode labelled as $u$ in Eq.~\eqref{eq:expansion2} represents the propagation of waves outwards the event horizon. Analogously to Eq.~\eqref{vac}, the outgoing vacuum state is defined as
\beq
a^{v, out}_{\omega}|0,out\rangle = 0,\,\,\,a^{u r, out}_{\omega}|0,out\rangle = 0,\,\,\,a^{u l, out}_{\omega}|0,out\rangle = 0.
\eeq
A Bogolyubov relation connects the annihilation operators in the incoming basis to those in the outcoming basis,
\bea
\left(\begin{array}{c}
  a^{v, out}_{\omega} \\
  a^{u r, out}_{\omega} \\
  (a^{u l, out}_{\omega})^{\dagger} 
\end{array}\right)
&=&
S
\left(\begin{array}{c}
  a^{v, in}_{\omega} \\
  a^{3, in}_{\omega}  \\
  (a^{4, in}_{\omega})^{\dagger}
\end{array}\right),
\label{bogo}
\eea
encoded in the unitary matrix $S$
\bea
S=\left(\begin{array}{ccc}
    S_{vl,vr} & S_{vl,3l} & S_{vl,4l} \\
    S_{ur,vr} & S_{ur,3l}  & S_{ur,4l} \\
    S_{ul,vr} & S_{ul,3l}  & S_{ul,4l}	  
\end{array}\right).
\eea
The physical interpretation is that the matrix $S$ encodes the mixed correlation among pair of modes. Indeed, the matrix element $S_{ul,vr}$, for instance, indicates that the incoming channel (second index) is a $v$-mode entering from the right region, while the outgoing channel (first index) is a $u$-mode escaping in the left region.  All the relevant expressions for the matrix elements of the Bogolyubov matrix~\eqref{bogo}  are collected in Refs.~\cite{may11}.

%%%%%%%%%%%%%%%%%%%%
\section{Results and discussion}

\begin{figure}
 \centering
    \includegraphics[width=8.8cm]{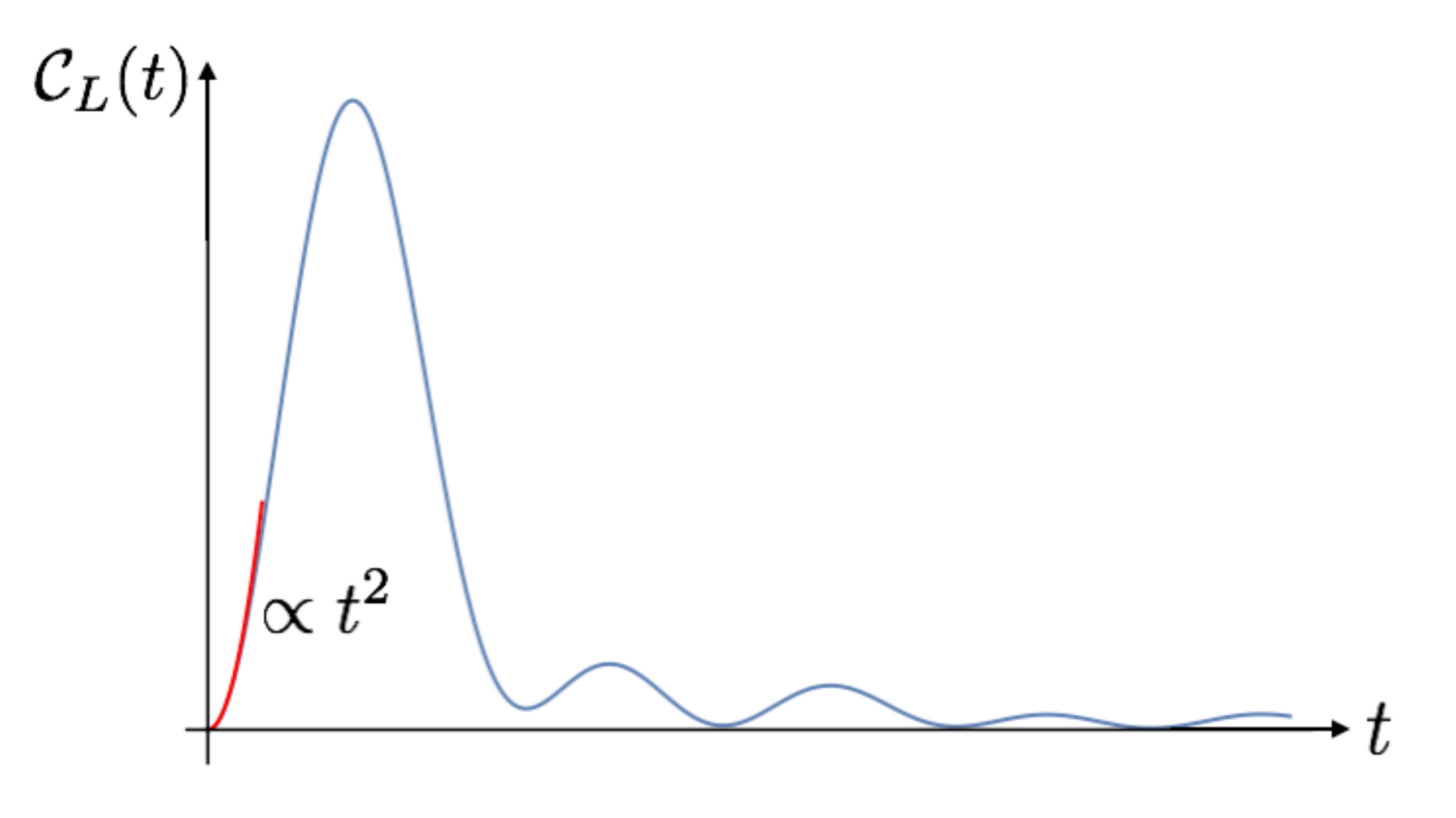}
    \caption{Polynomial growth $\propto t^2$ of the out-of-time order correlator  of the density of the Bose gas, ${\cal C}_{L}(t)$, at short times, indicating the slow scrambling character of the sonic black hole.
    At later times, $t\gg1/\Omega$, ${\cal C}_{L}(t)$ decays proportionally to $1/t^2$.
 }
     \label{fig2}
\end{figure}

After this concise review of the steps required to reproduce our computations, we now present our main findings. We calculate the out-of-time order commutator
\beq
{\cal C}_{L}(t) =   \langle in | [\delta\rho(t,x) , \delta\rho(0,x)]^{\dagger} 
[\delta\rho(t,x),\delta\rho(0,x)] | in \rangle,
\label{corr_comm_L}
\eeq
with respect to the $| in \rangle$ vacuum state and at a spatial point in the interior of the dumb hole. {As mentioned in the Introduction, out-of-time order commutators and correlators are related via Eq.~\eqref{relaz}.} In order to calculate~\eqref{corr_comm_L} we employ the expansion~(\ref{eq:expansion2}) for $\delta\psi(t,x)$ and use the relation~\eqref{bogo} in order to convert outgoing modes into incoming ones. In general, all the correlations among right and left modes contribute to the out-of-time order commutator, and self-correlations of left and right modes are also present. Nevertheless, we choose to focus only on the correlations of the modes  that would constitute the main signal of the Hawking radiation~\cite{may11}. The OTOC in Eq.~\eqref{corr_comm_L} then yields 
\bea
\label{eq:corr}
{\cal C}_{L}(t) &=& 4{\cal S}_{L}\left(\frac{(\sin (t \Omega )-t \Omega  \cos (t \Omega ))^2}{t^4}\right)
\nn\\
&+&\, 4{\cal T}_{L}\left(\frac{\sin ^4\left(\frac{t \Omega }{2}\right)}{t^2}\right),
\eea
where the coefficients $S_L$ and $T_L$ read
\bea
{\cal S}_{L} &=& \frac{\rho_{0}^{2}}{16 \pi^{2} m^{2} }\frac{c_{r}^{2}}{c_{l}^{4}(c_{r} - v_{0})^{4}},
\nn\\
{\cal T}_{L} &=& \frac{\rho_{0}^{2}}{8\pi^{2}}\frac{\left(v_{0} - c_{l}\right)^{3} (c_{l} + c_{r})^{2} (c_{r} + v_{0})^{2}}{c_{l}^{4}  (c_{l} - c_{r})^{2} (c_{r} - v_{0})^{2}(c_{l}+v_{0})}.
\eea
Analogue expressions hold for the out-of-time order commutator in the upstream region of the sonic black hole. For times shorter than $1/\Omega$, the out-of-time order commutator  ${\cal C}_{L}(t)$ scales algebraically  $\propto t^2$, while it decays and saturates (as also reported in other studies on scrambling~\cite{kitaev,Das, ON, bagrets, dicke,moess}) on longer time scales, $t\Omega\gg1$, as $1/t^2$; see for a summary Fig.~\ref{fig2}. The initial slow scrambling behaviour of the sonic black hole is consistent with the analogue character of OTOCs found in Luttinger Liquids~\cite{moess}, which grow with the same algebraic character at short times, indicating that a large class of dumb holes, in sharp contrast with  their astrophysical counterparts, is a slow scrambler of the quantum  information encoded in their operators.

Within the same considerations as above, we find that the out-of-time-order commutator for points on the opposite side of the sonic horizon~\footnote{{Eq. (14) is a special case of Eq. (11) when $x\to0$. However, since there is a discontinuity at this point, signaled by the presence of the event horizon of the sonic black hole, one should consider the limit $x\to0$ with due care. We have chosen to include the investigation of the hybrid situation in which the out-of-time order commutator is calculated for points on the opposite side of the horizon, before taking the limit $x$, $x' \to 0$; in this way, one can probe a regime which is naturally inaccessible for observers outside an astrophysical black holes. However, we have found that the algebraic growth of the OTOC $\propto t^2$ does not depend on the details of the choice of $x$ and $x'$.}}, defined by
\beq
\hspace{-1mm}
{\cal C}(t) =   \lim_ {x, x' \to 0} \langle in | [\delta\rho(t,x) , \delta\rho(0,x)]^{\dagger} 
[\delta\rho(t,x'),\delta\rho(0,x')] | in \rangle,
\label{corr_comm}
\eeq
has a form identical  to Eq.~\eqref{eq:corr} (at least with respect to the  contributions due to the Hawking signal), with the coefficients ${\cal S}_{L}$ and  ${\cal T}_{L}$, replaced, respectively, by
\bea
{\cal S} &=& \frac{\rho_{0}^{2}}{16 \pi^{2} m^{2} c_{l}^{2}(c_{r} - v_{0})^{4}},
\nn\\
{\cal T} &=& \frac{\rho_{0}^{2}}{2\pi^2}\frac{\left(v_{0} - c_{l}\right)^{3}(c_{l}+v_{0})}
{ c_{l}^{2} (c_{l}-c_{r})^{2} (c_{r}-v_{0})^{2}}.
\eea
\textcolor{red}{.}
The slow scrambling character of the type of dumb holes considered here can be interpreted as a result of the Bogolyubov treatment of weakly interacting bosonic gases, which reduces the original interacting problem to an effective quadratic quantum system. Indeed, fast scrambling is crucially connected with the nature of interactions in the system. On the other hand, one could argue prematurely that the outcome discussed above is instead a consequence of the usage of step-like sonic horizon models. Indeed, it is known that within such an idealized setting it is not possible to properly define a temperature~\cite{book1}. However, such assertions are completely unwarranted: Studies with more realistic horizon profiles~\cite{larre} still fully rely on a Bogolyubov expansion. Hence the algebraic growth of the OTOC presented in Eq.~\eqref{eq:corr} is still expected to emerge within such more realistic scenarios. This is a direct consequence of the fact that quadratic models in condensed-matter settings cannot host fast scrambling (as pointed out in Ref.~\cite{moess} in the context of Luttinger Liquids). Therefore a different shape of the sonic horizon will only change non-universal prefactors in Eq.~\eqref{eq:corr}, but not the character of the initial time growth of the OTOC. Furthermore, the gravitational analogy is strictly speaking confined to the hydrodynamical approximation in Bose-Einstein condensates~\cite{book1}. In particular, the long-standing problem of the possible information loss during black-hole evaporation is not present in the context of acoustic horizons, corroborating our finding that the sonic and astrophysics horizon cannot belong to the same class of quantum information scramblers.

Measurements of out-of-time order correlators are a challenge for experiments, and only a few proposals of  schemes involving Ramsey interferometry~\cite{yao} or echo-type protocols~\cite{exp}, have been currently scrutinised. A first result in this direction has been achieved implementing  time reversal of the many-body dynamics~\cite{gart} in a system of  trapped ions that simulate a collective interacting spin model. In this respect, one of the advantages of our work is the potential experimental accessibility of OTOCs of the type~\eqref{corr_comm_L}: since high-order correlators satisfy Wick's theorem when interactions are treated in Bogolyubov  approximation, it is possible to benefit from the two-point function measurements recently achieved in low-temperature and low-density dumb holes~\cite{expBH2}, in order to re-construct the slow scrambling character at short times of the density-OTOCs derived here.

%%%%%%%%%%%%%%%%%%%%
\section{Conclusions}

A Bogolyubov  theory of quantum fluctuations in a sonic black hole is sufficient to derive  an analogue Hawking radiation, and, accordingly, our computations for the OTOCs have been carried out at the same level of approximation in order to demonstrate that emission of pairs of particles close to the horizon is accompanied by slow scrambling, at least for a large class of sonic black holes. A natural extension would consist of going beyond this gaussian expansion and checking the fate of the $\propto t^2$ growth found in this work;  analogue numerical studies of the scrambling properties of Luttinger Liquids have, however, not shown remarkable discrepancies~\cite{moess}. From a broader perspective, this work might also represent a first step towards the study of out-of-time order correlators  in analogue models of gravity~\cite{liberati}, with the perspective to improve the understanding on quantum information scrambling from models  with much easier tunabilty and control  than their astrophysical counterparts. In the long run, this program could shed new  light on a  recent ongoing connection among the physics of scrambling and hydrodynamics instabilities~\cite{blake, blake2, grod}.

\acknowledgments
We  thank I. Carusotto for clarifying discussions on the physics of dumb holes; we are also grateful to  M. J. Holland and A. M. Rey for proof reading the manuscript and providing useful comments.

\bibliography{biblioSH}{}
\bibliographystyle{eplbib}

\end{document}